\documentclass[fleqn,usenatbib]{mnras}

\usepackage{newtxtext,newtxmath}

\usepackage[T1]{fontenc}

\DeclareRobustCommand{\VAN}[3]{#2}
\let\VANthebibliography\thebibliography
\def\thebibliography{\DeclareRobustCommand{\VAN}[3]{##3}\VANthebibliography}

\usepackage{graphicx}	
\usepackage{amsmath}	
\usepackage{tabularx}

\title{Interplanetary scintillation and pulsar pulse statistics}

\author[IPS and pulse statistics]
{ S. A. Tyul'bashev,$^{1}$ \thanks{E-mail: serg@prao.ru}
I. V. Chashei $^{1}$, 
	M. A. Kitaeva,$^{1}$
	\\
	$^{a}$ Lebedev Physical Institute, Astro Space Center, Pushchino Radio Astronomy Observatory,   Russian Academy of Sciences, Pushchino, 142290 Russia\\  
}

\date{December 22, 2022}

\pubyear{December 22, 2022}

\begin{document}
	\label{firstpage}
	\pagerange{\pageref{firstpage}--\pageref{lastpage}}
	\maketitle
	
	\begin{abstract}
The effect of interplanetary plasma on pulsed pulsar radiation passing through is considered. The pulses of two rotating radio transients (J0609+16, J1132+25) and a pulsar (B0320+39) detected on the Large Phased Array (Pushchino observatory) were analyzed. It is shown that in observations at the frequency of 111 MHz, on elongations of $20\degr-40\degr$, both an increase and a decrease in the number of received pulses are observed. The change in the number of pulses is explained by the distortion of the energy distribution of pulses due to interplanetary scintillation. These changes in the number of observed pulses are in qualitative agreement with the expected dependence of the scintillation index on the observed sources elongation. Analytical expressions are obtained that allow estimating the effective modulation index from observations of individual pulses for the power distribution of pulses by energy. 	
	\end{abstract}
	
	
	
	
	\section{Introduction}
    \label{intro}

As is known, pulsars were discovered during the research of interplanetary plasma by observing a scintillation of compact ($<1-2\arcsec$) radio sources (\citeauthor{Hewish1968}, \citeyear{Hewish1968}). 
Despite the discovery of pulsars actually by the method of interplanetary scintillation in the study of the interplanetary medium, pulsars did not make a special contribution to its study. This is due to the fact that, in general, an integral flux density of a pulsar is small and for observations of pulsars as compact radio sources of constant radiation that shine through interplanetary plasma, antennas with high sensitivity are needed. Thus, in the catalog of scintillation radio sources of the northern hemisphere (\citeauthor{Purvis1987}, \citeyear{Purvis1987}), obtained at a frequency of 81.5 MHz, out of 1,789 compact sources in the catalog, only 17 are identified as pulsars. When studying the interplanetary medium, it is much more profitable to use a core of a quasar or a radio galaxy as a source that shines through plasma on a line of sight, rather than a pulsar. 

There are very few early works with the special use of pulsars for interplanetary plasma research. We note only one of them, which uses pulsars as sources of very small angular sizes. \citeauthor{Zeissig1972} (\citeyear{Zeissig1972}) provides observations of interplanetary scintillation of pulsars CP   0950 (J0953+0755) and NP 0527 (J0528+2200) at a frequency of 318 MHz. For the pulsar J0953+0755, which has a dispersion measure $DM \approx 3$ pc/cm$^3$, the interstellar medium practically does not scatter pulses, and the index of scintillation ($m_0 (\epsilon)$), depending on the elongation, ($\epsilon$ is the angle between the direction to the Sun and to the source when observed from the Earth) reaches a maximum and remains constant (see Fig. 3 in \citeauthor{Zeissig1972} (\citeyear{Zeissig1972})). For the pulsar J0528+2200 $DM \approx 51$ pc/cm$^3$, the interstellar medium scatters pulses, making a pulsar a non-point object for an observer. The dependence $m_0 (\epsilon)$ reaches a maximum, and then falls to small elongations (see Fig. 4 in \citeauthor{Zeissig1972} (\citeyear{Zeissig1972})). Numerical calculations for the interplanetary scintillation show exactly this behavior of $m_0 (\epsilon)$ for point and non-point sources (\citeauthor{Marians1975}, \citeyear{Marians1975}). For a point source the scintillation index at the maximum (focusing zone) may exceed unity (\citeauthor{Prokhorov1975}, \citeyear{Prokhorov1975}).

\begin{figure*}
	\includegraphics[width=0.7\textwidth]{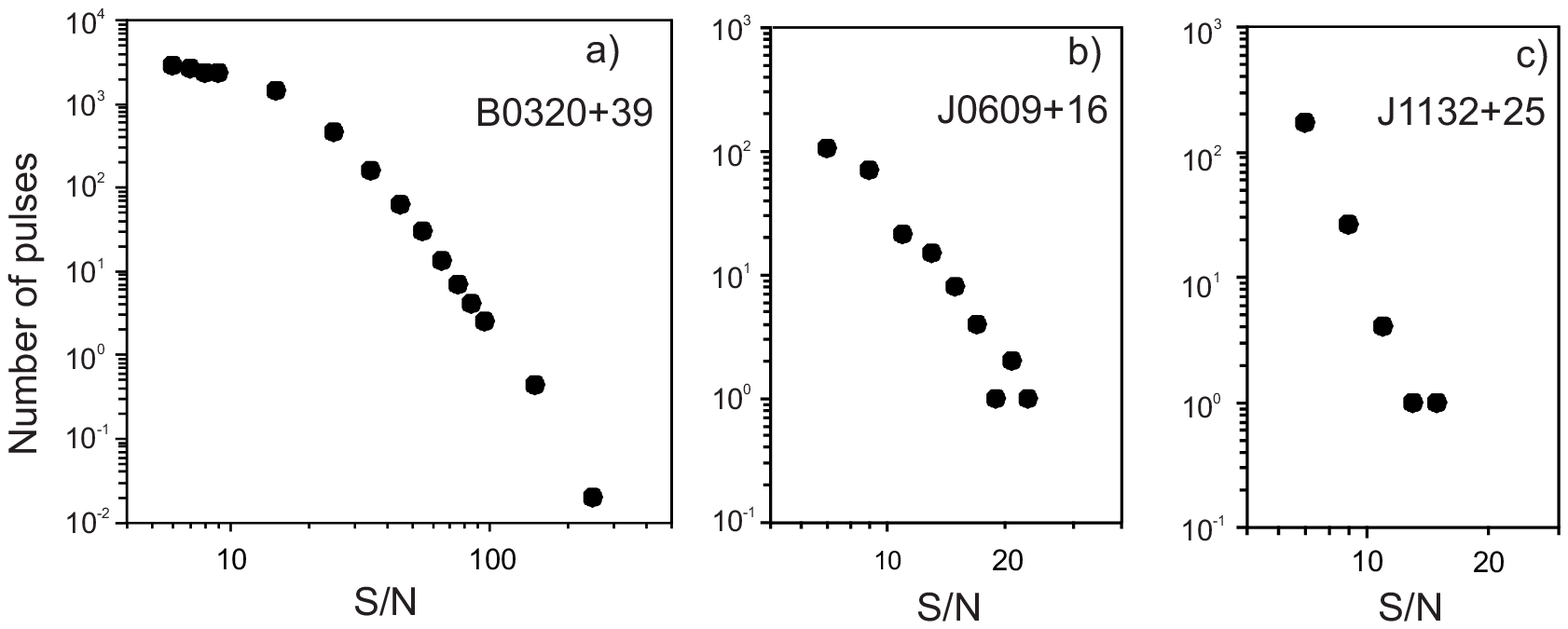}
	\caption{Energy distribution of pulses for pulsar B0320+39 and for two RRATs (J0609+16; J1132+25). The vertical axis shows the number of pulses found, the horizontal axis shows the S/N for these pulses.}
	\label{fig1}
\end{figure*}

In recent years, works have appeared (\citeauthor{Tyulbashev2018}, \citeyear{Tyulbashev2018}, \citeauthor{Smirnova2022}, \citeyear{Smirnova2022}) indicating that it is possible to use interplanetary plasma as an amplifier of signals from rotating radio transients (RRAT). RRAT, being a subsample of pulsars, were discovered as sources of pulsed radiation having a dispersion delay of the signal due to its passage through the interstellar medium (\citeauthor{McLaughlin2006}, \citeyear{McLaughlin2006}). A pulse radio emission is not detected in the majority of pulsar rotations. In \citeauthor{Tyulbashev2018} (\citeyear{Tyulbashev2018}) it was shown that the interplanetary medium can amplify signals passing through it by 1.5-2 times. The estimated number of RRATs in the Galaxy is twice as large as the number of ordinary pulsars (\citeauthor{Keane2008}, \citeyear{Keane2008}). In the ATNF pulsar catalog\footnote{https://www.atnf.csiro.au/research/pulsar/psrcat/} (\citeauthor{Manchester2005}, \citeyear{Manchester2005}) out of more than 3,000 pulsars, there are just, approximately, one hundred  rotating radio transients. Obviously, the shortage of undetected RRATs is very large. For weak RRATs, whose pulses are beyond the detection threshold in the meter radio range, signal amplification, leading to an increase in a pulse amplitude, may be the only way to discover them.

In \citeauthor{Smirnova2022} (\citeyear{Smirnova2022}), 16 RRAT pulses were studied at a frequency of 111 MHz and it was found that for six of them, the number of observed pulses $N (\epsilon)$ increases from 2 to 10 times by $\epsilon = 30\degr \pm 10\degr$. When studying this effect, it was proved that an increase in the number of observed pulses is associated with an increase in weak pulse signals on the interplanetary plasma. 

However, there are three different manifestations of interplanetary scintillations associated with a passage of pulsar pulses through the interplanetary plasma. Some part of undetectable pulses, the signal from which is below the detection threshold, are amplified by scintillation and become observable. As noted above, this effect was published in the paper \citeauthor{Smirnova2022} (\citeyear{Smirnova2022}), but its detailed study was not carried out in this work. Secondly, if there is an increase or decrease in the energy of individual pulses associated with the interplanetary medium, we can talk about the scintillation index, reflecting these changes in numerical form. A developed method for estimating $m_0 (\epsilon)$ from observations of individual pulsar pulses does not exist to date. The measurements of $m_0 (\epsilon)$ by individual pulses will differ from the measurements in the work of \citeauthor{Zeissig1972} (\citeyear{Zeissig1972}), where a pulsar was considered as a compact radio source with constant radio emission. Thirdly, if the number of pulses with different energies changes due to interplanetary scintillation, a distortion of the dependence of the pulsar pulse distribution by the energies should appear.  We consider all three effects in this paper.

\section{Observations}
\label{Observation}

RRATs emits their pulses sporadically and hours may pass between subsequent pulses. Therefore, it takes a lot of time to detect the number of pulses needed for analysis. RRAT and pulsar observations were carried out on the Large Phased Array (LPA). LPA antenna is a meridian instrument, and the transit time through the meridian takes about 3.5 minutes per day at half power. A daily monitoring survey (Pushchino Multibeams Pulsar Survey - PUMPS (\citeauthor{Tyulbashev2022}, \citeyear{Tyulbashev2022}) has been carried out on the antenna for 8 years. LPA beams covers approximately 17,000 square degrees of the northern hemisphere every day with 96 beams.

In this paper, we use the signal to noise ratio (S/N) pulses of two RRATs (J0609+16; J1132+25) and one pulsar (B0320+39), previously detected during the study of these objects (\citeauthor{Smirnova2022}, \citeyear{Smirnova2022}, \citeauthor{Brylyakova2021}, \citeyear{Brylyakova2021}). Observations of the sources were carried out at the central frequency of 110.3 MHz, in the 2.5 MHz band, with a 12.5 ms sampling and using 32 frequency channels in band.

\section{Influence of the interplanetary medium on the observed pulses}
\label{Influence}

Let us consider effects of the passage of pulsed radiation through the interplanetary medium, indicated in the Section~\ref{intro}, from a physical point of view. The number of observed RRAT pulses according to \citeauthor{Smirnova2022} (\citeyear{Smirnova2022}) increased at elongations of $20\degr - 40\degr$ (optimal elongations at meter wave length). These elongations correspond to the maximum in the dependence $m_0 (\epsilon)$ for conventional compact radio sources. Our reasoning in explaining the increase in the number of observed pulses was as follows. For pulsars, only the strongest pulses are usually observed. The weaker are the pulsar pulses, the more of them can be observed. Interplanetary scintillation amplifies and attenuates pulses that are above and below their detection threshold. Increasing the number of detections of undetectable pulses due to the scintillation should be observed if the pulse energy distribution is decreasing function. This will lead to an increase in the number of detected pulses in general. However, the situation is not as simple as it seems. Even for the RRATs studied by us, a noticeable increase in the number of detected pulses was observed only for 6 of the 16 studied ones (\citeauthor{Smirnova2022}, \citeyear{Smirnova2022}). The energy distribution of pulses in pulsars most often obeys the lognormal law, sometimes a power tail is observed (see, for example, \citeauthor{Mickaliger2018} (\citeyear{Mickaliger2018}). The lognormal distribution of B0320+39 can be conditionally divided into three sections (Fig.\ref{fig1}a), the boundaries of which for each pulsar will be determined by different signal-to-noise ratios (S/N). J0609+16 and J1132+25 have a power distribution with spectral indices of -4 and -6.5 (Fig.\ref{fig1}b,c).

\begin{figure*}
	\includegraphics[width=0.8\textwidth]{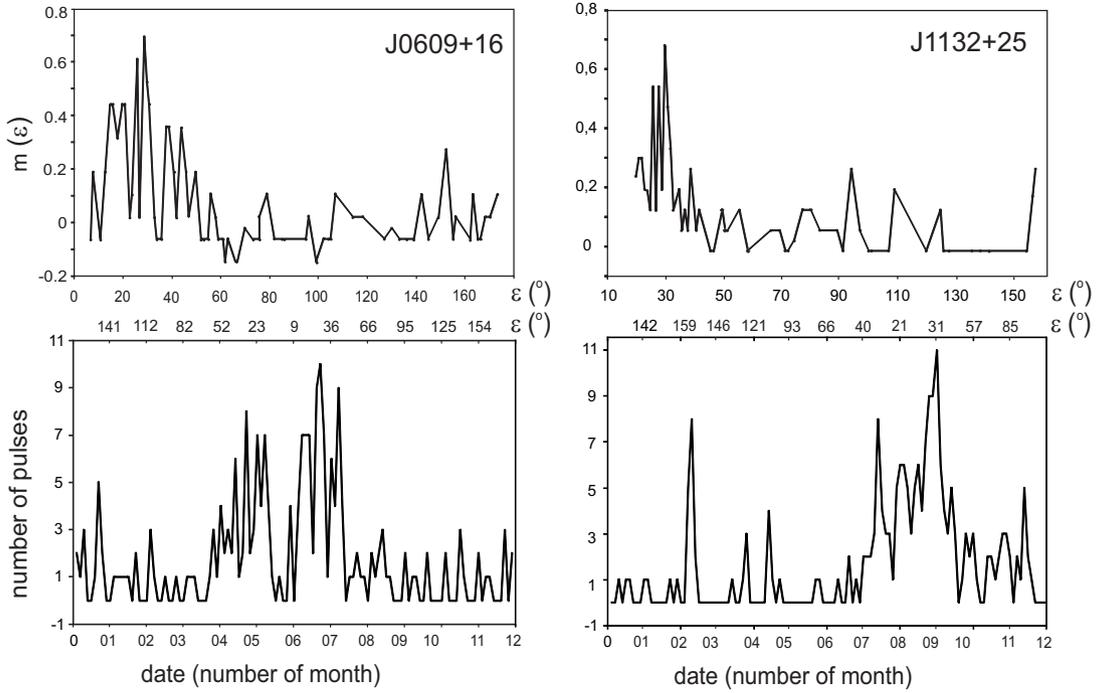}
	\caption{The left and right panels show the dependencies $m(\epsilon)$ for RRATs J0609+16 and J1132+25 from $\epsilon$ and number of detected pulses in different months. The values of $m(\epsilon)$ were estimated with eq.~\ref{eq10} according to data from \citeauthor{Brylyakova2021} (\citeyear{Brylyakova2021}), \citeauthor{Smirnova2022} (\citeyear{Smirnova2022}).}
	\label{fig2}
\end{figure*}

It is obvious that for different sections of the energy distribution of B0320+39 pulses, the effects indicated in the Introduction will manifest themselves in different ways. It is qualitatively clear that in the region S/N $ > 20$ the distribution should flatten. There should be no changes at the top of the distribution (on a flat range). In the region S/N $ < 7$, there should be an increase of steepness of the distribution.

The balance of the observed number of pulses when transferring pulses from undetectable to detectable and vice versa depends on the type of pulse energy distribution. In the case of a power law distribution with power-law index $\alpha >1$, the total number of observed pulses will be higher than it would be in the absence of an interplanetary medium. If the distribution is lognormal and a peak is visible in it, then in the presence of an interplanetary medium, the balance should be reversed. That is, the total observed number of pulses will be less than in the absence of interplanetary plasma. However, whatever a balance, the observed energy distribution of pulses should be distorted.

Since pulsars are generally weak objects and individual pulses are observed in a very small part of them, most often a peak in the distribution is not observed. In this case, there will be more undetectable pulses than detectable ones, even for a lognormal distribution. Therefore, the transfer of undetectable pulses into detectable pulses should lead to an overall increase in the number of observed pulses. To obtain a correct estimate of the number of\, ''additional``\, pulses acquired or lost, it is necessary to know the energy distribution of the pulses and the sensitivity limit of the radio telescope. As shown in \citeauthor{Smirnova2022} (\citeyear{Smirnova2022}), the contribution of diffraction (short in time) scintillation to the change in the pulse flux density for the studied RRAT is absent due to strong smoothing in wide frequency channels. 

The typical time of refractive scintillation in the meter wavelength range is tens and hundreds of days. At the interval of pulsar recording  on the LPA LPI of a few minutes, refractive scintillation will equally strengthen or weaken all the observed pulses. Thus, the main changes will be associated with their own variations in the pulse flux density. These variations are reflected in the energy distribution of pulses and should be taken into account when evaluating the scintillation index. 

Above in this paper, we qualitatively considered the effects associated with the influence of interplanetary plasma on the observed pulse energy. Below we get analytical expressions that allow us to quantify these effects.

Let $f(I)$ be the differential and $F(I)$ be the integral functions of the pulsar distribution over the pulse intensity $I$,
\begin{equation}
 f (I) = \frac{dF (I)}{dI} 
	\label{eq1}
\end{equation}   
then the average intensity is
\begin{equation}
	\langle I \rangle = \int\limits_{I_{min}}^{I_{max}} I f(I) dI ,
	\label{eq2}
\end{equation}           
the total number of detected pulses  ($N_0$) without taking into account scintillation is
\begin{equation}
N_0	 = \int\limits_{I_{min}}^{I_{max}}  f(I) dI ,
	\label{eq3}
\end{equation}   
here $I_{max}$, $I_{min}$ are the maximum and minimum pulse intensities.


Interplanetary scintillation can change the shape of the pulse energy distribution $f(I)$ when the pulsar solar elongation angle is less than $90^o$. In this case distribution  $f''(I)$ is the convolution of intrinsic energy distribution of pulses with intensity scintillation distribution $\phi(I, I')$

\begin{equation}
f'(I)= \int \phi(I,I') f(I')dI'.
\label{eq11}
\end{equation}   

According to \citeauthor{Mercier1962} (\citeyear{Mercier1962}) scintillation statistics is described by Rice distribution. In the central part this function is close to symmetric one relative to mean intensity $I'$ in the regime of weak scintillation, when the scintillation index $m_0 < 1$. Scintillation intensity is distributed by nonsymmetrical exponential function in the regime of saturated scintillation, when $m_0 = 1$. The intensity statistics is dominated by the values of $I$ less than mean value $I'$ (see, for instance, \citeauthor{Morgan2018} (\citeyear{Morgan2018}). We cannot find universal analytical expression for $f'(I)$ even for simplest, particularly flat,  $f(I)$ distributions, due to the complicated form of Rice distribution. For this reason, to study the influence of scintillation on pulse statistic we propose here the simple approach assuming that scintillation intensity is uniformly distributed in the limits $(1-m)I < I < (1+m)I$. We consider this flat intensity distribution as equivalent to real convolution. We can present two arguments in favor of such approach: i) weak scintillation regime occupies main duration within the year observation series; ii) in contrast with interstellar scintillation, when all the observations are in strong scintillation, only small parts of series are related to strong scintillation regime because approaching the Sun the elongation angle decreases for about 1$^o$ during a day due to the source motion relative to the Sun. The value of introduced effective parameter $m$ is not equal to classical scintillation index $m_0$ , but we can expect an approximate proportionality between two values. In the frame of our assumption the change in the number of pulses due to scintillation will be equal to
\begin{equation}
\Delta N (I) = - F [  (1+m) I] - F [  (1-m) I] , 
	\label{eq4}
\end{equation} 
which at $m < 1$ is approximately equal to
\begin{equation}
	\Delta N (I) \approx - 2 m I f(I)  . 
	\label{eq5}
\end{equation} 

\begin{figure*}
	\includegraphics[width=0.8\textwidth]{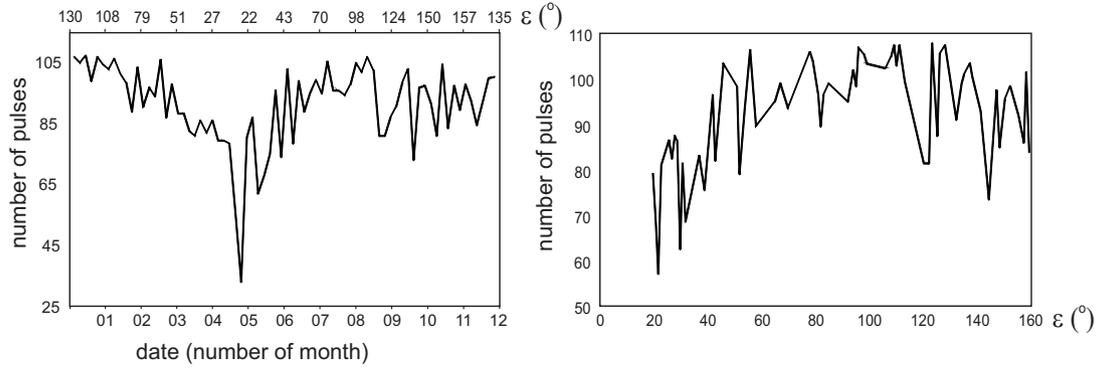}
	\caption{The left panel shows the number of observed pulses (vertical axis) from the date (horizontal axis) for the pulsar B0320+39. The date is shown as the month number. The right panel shows the number of observed pulses (vertical axis) depending on the elongation (horizontal axis).}
	\label{fig3}
\end{figure*}

The change in the distribution function associated with scintillation will be equal to
\begin{equation}
\Delta f = \frac{d \Delta N (I)}{dI} \approx - \frac{2 m d [I f (I)]}{dI},
	\label{eq6}
\end{equation}   
from where we find the complete change in the number of pulses registered in the session
\begin{equation}
\Delta N \approx  -2  m  [ I_{max} f (I_{max}) -  I_{min} f (I_{min}) ], 
	\label{eq7}
\end{equation} 
 where $I_{max}$ and $I_{min}$ are the values of the maximum and minimum pulse intensities depending on the experimental conditions. The value of $\Delta N$, as follows from eq.~\ref{eq7}, significantly depends on the shape of the distribution function $f(I)$. In the case of a monotone-growing distribution, the observed number of pulses will decrease due to scintillation. In the case of a monotone-decreasing distribution, scintillation leads to an increase in the number of pulses. For a non-monotone distribution, depending on the specific type of $f(I)$, both an increase and a decrease in the number of pulses can be observed. In this case, the change in the number of pulses, which is proportional to the scintillation index, will depend on the angular distance between the pulsar and the Sun.

For 6 RRATs from \citeauthor{Smirnova2022} (\citeyear{Smirnova2022}), the distribution function has a decreasing power-law form
\begin{equation}
f(I) = A \left( \frac{I_{min}}{I} \right)^{-\alpha} , 
	\label{eq8}
\end{equation} 
where $A = const$. In this case, assuming $I_{max} \gg I_{min}$, we find from eqs.\ref{eq7}, \ref{eq8} the total observed number of pulses
\begin{equation}
N_{obs} = N_0 + \Delta N = N_0 [1 + 2m (\alpha - 1)], 
	\label{eq9}
\end{equation} 

\begin{equation}
m = \frac{N_{obs} - N_0}{2 (\alpha - 1) N_0}.     
	\label{eq10}
\end{equation} 

For pulsars, the distribution function is usually nonmonotonic and is lognormal or lognormal with a power law\, ''tail``. In this case, the ratio between $N_{obs}$  and $m$, as follows from eq.~\ref{eq7}, will be more complex than eq.~\ref{eq9}, since it is determined by the competition between pulses of increased and decreased due to intensity scintillation. The ratios eqs.~\ref{eq3}, \ref{eq4}, \ref{eq6}, \ref{eq7} allow us to estimate the scintillation index in the case of an arbitrary distribution of pulse intensity, including with a decrease in their observed number. At the same time, however, it was not possible to obtain an analytical relation similar to eqs.~\ref{eq9}, \ref{eq10}.

The obtained ratios confirm the qualitative considerations given at the beginning of the paragraph.

\section{Analysis of observation data}

To estimate the scintillation index, two RRATs (J0609+16; J1132+25) from \citeauthor{Smirnova2022} (\citeyear{Smirnova2022}) were selected, showing a noticeable increase in the number of observed pulses at elongations optimal for observations in the meter wavelength range. Eq.~\ref{eq10} was used to calculate the scintillation index. The average number of observed pulses outside the zone of optimal elongations was taken as $N_0$ value. The energy distribution of pulses for these RRATs turned out to be power-law and the observed section in the energy distribution of pulses on a logarithmic scale is a straight line, and the spectral index $\alpha$ determines the inclination angle of this straight line. The paper \citeauthor{Smirnova2022} (\citeyear{Smirnova2022}) provides a link to tables containing the dates of observations and S/N for observed pulses of the tested RRATs. We have calculated the scintillation indices, and the developed dependencies $m(\epsilon)$ are shown in Fig.~\ref{fig2}.

Fig.~\ref{fig2}. shows a strong roughness depending on $m (\epsilon)$ at elongations $\epsilon = 30\degr \pm 10\degr$. We assume that it is due to the fact that the number of observed pulses, including at optimal elongations, is low. Therefore, even small changes in the number of observed pulses produce large errors depending on $m(\epsilon)$.

As shown in the previous paragraph, in the region of small S/N (see Fig.~\ref{fig1}a) at optimal elongations for log-normal pulse distribution, detectable pulses should be transferred into undetectable ones, thereby reducing the total number of pulses. Fig.~\ref{fig3} illustrates this behavior for the strong pulsar B0320+39. The drop in the observed number of pulses in March-June due to interplanetary scintillation for this pulsar is due to the fact that the main contribution to the variations is made by the growing portion of the pulse energy distribution (left section on the Fig.~\ref{fig1}a). To smooth out the dependence on the left panel of Fig.~\ref{fig3}, the addition of the observed number of pulses was carried out for intervals of three days. When constructing the dependence on the right panel, when the elongations coincide before and after passing the minimum elongations, averaging was carried out.

The figure on the left panel shows that the average number of pulses B0320+39 is approximately 100 and falls 2-3 times in April-June. In May, the pulsar passes through $\epsilon = 20\degr - 25\degr$ and the number of observed pulses drops below 50. From the drawing on the right panel it can be seen that the minimum number of pulses is detected at small elongations. To obtain the dependence, the number of detected pulses was averaged before and after the passage of the minimum elongation observed for this pulsar. For the lognormal distribution, an analytical expression describing the behavior of the pulse energy distribution function at different sites could not be obtained, and therefore the scintillation index for B0320+39 was not calculated.

\section{Discussion of the results}

The paper considers three manifestations of the interplanetary scintillation on the observed pulse statistics of pulsars: a change in the number of observed pulses, a distortion of the number of pulses distribution by energy, the appearance of the scintillation index dependence on the observed elongation.

For 6 transients (\citeauthor{Smirnova2022}, \citeyear{Smirnova2022}), an increase in the number of observed pulses from 2 to 10 times by $\epsilon \approx 30\degr \pm 10\degr$ was previously shown. That is, it is shown that the search for transients should be carried out at the elongations optimal for the meter wavelength range. In the present work, for the pulsar B0320+39, it is shown that at the same elongations, a drop in the number of observed pulses by several times can be observed. As shown above, an increase or decrease in the number of detected pulses depends on the energy distribution of the pulses. If the weakest observed pulses in the distribution fall on the monotone falling part of the dependence, then the number of observed pulses will increase. If a maximum is visible in the distribution, and then a drop in the number of detected pulses is observed, then less pulses will be detected at $\epsilon \approx 30\degr \pm 10\degr$ than outside this elongation zone.

The considered effect will be stronger for weak pulsars, in which many individual pulses are not visible. If a pulsar is strong and it has all or almost all pulses, the interplanetary medium will distort the distribution of pulses by energy, but will not add new pulses. In the case of a lognormal energy distribution of pulses, the right part of the distribution (the area containing strong pulses) is flattened, the left (the area containing weak pulses) becomes steeper. Therefore, it is incorrect to compare distributions obtained at different frequencies. For a correct comparison of distributions, it is necessary to preliminary delete observations with dates on which optimal elongations fall. For the observations considered here, these are the dates corresponding to $\epsilon = 20\degr - 40\degr$.


Classical scintillation index $m_0$ reaches maximal value about 1 at solar elongation about $\epsilon_m \simeq 25^o$ for our frequency 111 MHz. It decreases at the elongations less than $\epsilon_m$ due to finite source visible angular size that is defined by interstellar scattering. The interstellar scattering results also in pulse broadening. The value of $\epsilon_m$ is less than 25$^o$ at higher observation frequencies. The decrease of scintillation index in the range $\epsilon < \epsilon_m$ will be not so strong in this case because $sin \epsilon_m$ is approximately proportional to observation wavelength $\lambda$ while the scattering angle $\theta_{sc}$ is approximately proportional to $\lambda^2$. However some additional decrease in scintillation level can take place due to scintillation decorrelation across the frequency band if the observation bandwidth will be sufficiently broad. The last effect is not important for the LPA frequency band at the elongations $\epsilon \ge \epsilon_m$. The increase of phase fluctuation results in strong scintillation in the range $\epsilon < \epsilon_m$. However the observed scintillation level is suppressed in the strong scattering regime due to the finite source angular size and to decorrelation in the frequency band. Strong scintillation regime occupies only comparatively small part of the whole length of observation series used in our studies. It should be noted that the individual pulses are observable only for comparatively close sources with dispersion measure $DM < 100$ pc/cm$^3$ at the LPA operating frequency.

According to the expressions obtained in the Section~\ref{Influence} the effective scintillation parameter $m$ can be estimated. If we consider pulsars as sources of zero angular size, then the maximum value of classical scintillation index $m_0$ should reach unity. As can be seen from Fig.~\ref{fig2}, for both RRATs, the parameter $m$ reaches only 0.7. The difference between the maximum value of parameter $m$ and unity may be explained by the following factors: parameter $m$ is not exactly the classical scintillation index; the real exponential scintillation intensity distribution can result in fewer detected pulses; finite source angular sizes can reduce as $m_0$ so $m$.        

\section*{Data availability}
The table with S/N of B0320+39 pulses are in http://prao.ru/online\%20data/onlinedata.html

\section*{Acknowledgements}

The study was carried out at the expense of a grant Russian Science Foundation 22-12-00236, https://rscf.ru/project/22-12-00236/. We express our gratitude to L.B. Potapova for the preparation of drawings.



\label{lastpage}
\end{document}